\documentclass[twocolumn,floatfix,showpacs,prb,preprintnumbers,amsmath,amssymb]{revtex4}
\usepackage{graphicx}
\usepackage{natbib}
\usepackage{dcolumn}% Align table columns on decimal point

\begin{document}
\title{Field-induced structural evolution in the spin-Peierls compound CuGeO$_3$: high-field ESR study.}
\author{S.A.~Zvyagin, J.~Krzystek}
\affiliation{National High Magnetic Field Laboratory, 1800 East
Paul Dirac Drive, Tallahassee, FL 32310}
\author{  P.H.M. van Loosdrecht}
\affiliation{Department of Physics, University of Groningen,
Nijenborgh 4, 9747 AG, Groningen, The Netherlands}
\author{G. Dhalenne and A. Revcolevschi}
\affiliation{Laboratoire de Physico-Chimie de l'Etat Solide,
Universit$\acute e$ de Paris-Sud, 91405 Orsay Cedex, France}

\begin{abstract}

The dimerized-incommensurate phase transition in the spin-Peierls
compound CuGeO$_3$ is probed using frequency-tunable
high-resolution electron spin resonance (ESR) technique, in
magnetic fields up to 17 T. A field-induced development of the
soliton-like incommensurate superstructure is clearly indicated as
a pronounced increase  of the magnon spin resonance linewidth
$\Delta B$, with a $\Delta B_{max}$ at $B_{c}\sim$ 13.8 T. The
anomaly is explained in terms of the magnon-soliton scattering,
and suggests that the soliton-like phase exists close to the
boundary of the dimerized-incommensurate phase transition. In
addition, magnetic excitation spectra in 0.8\% Si-doped CuGeO$_3$
are studied. Suppression of the $\Delta B$ anomaly observed in the
doped samples suggests a collapse of the long-range-ordered
soliton states upon doping, that is consistent with high-field
neutron experiments.

\end{abstract}
\pacs{75.30.Kz, 75.40.-s, 76.30.-v} \maketitle

        The discovery of a spin-Peierls transition in an inorganic
        compound CuGeO$_3$\cite{Hase} has stimulated significant interest in experimental and
        theoretical
        studies of low-dimensional materials. A lattice dimerization,
        which is one of the most characteristic features in the
        spin-Peierls transition, was found to take place below T$_{SP}\sim$ 14
        K. In the dimerized phase the ground state is a spin singlet, separated from the
        first excited triplet by an energy gap. Application of
        external magnetic field tends to suppress quantum
        fluctuations, and eventually collapses the energy gap.
        By increasing the magnetic field above the threshold field, $B_{DI}\sim$ 12.5 T,
        CuGeO$_3$ undergoes a transition from the dimerized spin-liquid
        commensurate to the incommensurate phase, where the periodicity
        of the spin-polarization and lattice deformation is incommensurate
        with crystallographic lattice parameters. The low-field incommensurate region
        can be described by a formation of a regular array of domain walls (solitons).
        If the concentration of solitons is high enough, interactions between them
        result in a long-range ordered soliton lattice, observed experimentally\cite{Kiryukhin}.
        A further increase in field  induces a plane-wave modulated (harmonic) incommensurate
        state\cite{Lorenz,Horvatic}, where the modulation phase is a linear
        function of the space coordinate in the direction of the
        modulation.

          The rich magnetic phase diagram of CuGeO$_3$ has been a subject
          of many intensive high-field  investigations.
          The high-frequency/field electron spin resonance (ESR) technique was employed
    for studying magnetic excitation spectra in CuGeO$_3$
    \cite{Brill,Nojiri2,Nojiri3,Yamamoto,Wpalme}.
    These investigations provide valuable information on the size of
    the energy gaps in CuGeO$_3$ in the dimerized phase, on the $g$-factors, and on the exchange coupling.
    Like nuclear magnetic resonance methods, high-resolution ESR is a very powerful tool to
 study
    local spin environments in solids. It was successfully used for
    investigating structural incommensurability in various materials\cite{Blink}.
    Since the dimerized-incommensurate phase transition in
    CuGeO$_3$ has a magnetic origin, probing magnetic excitations  in a broad
    range of magnetic fields (and frequencies) can provide important
    information on field-induced structural evolution of
    CuGeO$_3$. The main motivation of this investigation was to study peculiarities of the dimerized-incommensurate phase
    transition in CuGeO$_3$, using multifrequency high-field high-resolution
    ESR.

In this work, we present a systematic study of the ESR linewidth
(spin triplet excitations) obtained in pure and 0.8\%Si-doped
CuGeO$_3$ single crystals, in the quasi-continuously covered
frequency range of 175-510 GHz and magnetic fields up to 17 T. To
the best of our knowledge, this is the first high-resolution ESR
investigation of the commensurate-incommensurate phase transition
in CuGeO$_3$, which is not driven by temperature, but by the
magnetic field.

Experiments were performed using the high-field millimeter and
submillimeter wave spectroscopy facility at the National High
Magnetic Fields Laboratory, Tallahassee, FL. A key feature of the
facility is a set of
    easily-tunable millimeter and sub-millimeter wave radiation sources,
    Backward Wave Oscillators (BWOs), operating in the frequency
    range of 140-700 GHz ($\sim$4.6 - 23.3 cm$^{-1}$). BWOs are classic vacuum-tube microwave
    devices, which (unlike other sources of millimeter and submillimeter wave
radiation)  possess an important distinguishing characteristic:
they are tunable over a very wide frequency range - up to ±30\%
from their central frequency. Due to this important property,
high-field BWO ESR spectroscopy provides a remarkable possibility
to probe magnetic excitations in a broad,
quasi-continuously-covered range of frequencies and magnetic
fields\cite{Naumenko,Zvyagin3} (unlike conventional ESR methods,
which employ one constant frequency, or a set of frequencies). The
BWOs,  in combination with highly-homogeneous
    (12 ppm/cm DSV) magnetic field provided by a 25 T hysteresis-free resistive magnet, make the
    facility a very powerful tool for systematic high-resolution ESR investigations of field-induced phenomena in
    CuGeO$_3$ and other magnetic materials.

The spectrometer works in transmission mode and employs oversized
 cylindrical waveguides.  An extremely low-noise, wide frequency range, InSb hot electron bolometer,
operated at liquid-He temperature, serves as a detector.
 The spectrometer allows for experiments to be carried out over a range of temperatures from 1.5 to 300 K.
 The spectra are recorded while sweeping the magnetic field. Two kinds of signal modulation are possible. While
 modulation of the magnetic field gives a better signal-to-noise ratio for narrower lines, modulation of the
 radiation
 power using a chopper (optical modulation) allows a direct detection of the absorption/transmission and provides better sensitivity
 for broader resonance lines. The spectrometer operates in the Faraday or Voigt  geometry (propagation vector of the
 radiation parallel or perpendicular to the external magnetic field, respectively).

 In order to detect the real shape of the
absorption with a minimum of experimental error, optical
modulation of the radiation power was used in our experiments.
Pure and 0.8\% Si-doped CuGeO$_3$ single crystals with a typical
thickness of 0.2 mm were used. The experiment was performed in the
Faraday geometry with magnetic field applied in the direction of
the $a$-axis. In this work we focused on studying the
dimerized-incommensurate phase transition, and thus only results
obtained in fields up to 17 T are presented. ESR investigation of
the magnetic excitations in CuGeO$_3$ at higher fields (in the
plane-wave modulated phase) is beyond the current consideration
and will be reported elsewhere\cite{Zvyagin2}.

Before going ahead with experimental data, let us briefly
characterize low-energy spin excitations in CuGeO$_3$. Above
$T_{SP}$,  CuGeO$_3$ is in the commensurate phase, and can be
regarded as an $S$=1/2 uniform Heisenberg antiferromagnet, with a
gapless spin singlet ground state. Triplet excitations in this
phase can be described as massless domain wall-like $S$=1/2
fermion-type excitations, spinons. Below $T_{SP}$, CuGeO$_3$ is in
the dimerized phase; the ESR spectrum is basically formed by
transitions between the excited Zeeman split triplet states; these
massive boson-type excitations can be defined as magnons, and the
corresponding resonance - as  a magnon spin
resonance\cite{Boucher}. With dimerization the spinons are
confined into magnon excitations; as a result, the two-spinon
continuum in the dimerized phase is significantly modified.
Transitions from the ground states are normally forbidden in low
dimensional gapped spin systems. However, breaking translational
symmetry (due to the Dzyaloshinskii-Moriya interactions, or
staggered field effects, for instance) can allow ground state
excitations. These transitions occur at the center of the
Brillouin zone; the observation of these transitions using ESR
provides direct and accurate information on energy gaps in
CuGeO$_3$\cite{Brill,Nojiri3}. In the soliton-like incommensurate
phase there are two types of competing excitations. Magnetic
excitations within the spin-dimerized domains can be ascribed to
the magnon subsystem (magnons), while  soliton-type excitations
originate from transitions within the soliton subsystems. The
soliton subsystem appears to strongly contribute to the bulk
magnetization and the excitation spectrum of the CuGeO$_3$ in the
soliton-like phase\cite{Enderle}. Magnetic bound states, which are
a general feature of many low-dimensional spin systems (see for
instance\cite{Date,Zvyagin}), manifest themselves in CuGeO$_3$ in
the far-infrared region\cite{Els}, and can be an interesting
subject for high-frequency/field ESR studies.

\begin{figure}
\begin{center}
\vspace{2cm}
\includegraphics[width=0.45\textwidth]{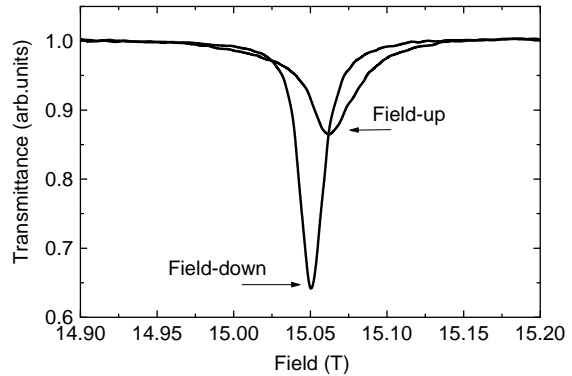}% Here is how to import EPS art
\vspace{-1.2cm} \caption{\label{Fig1} The ESR spectrum in
CuGeO$_3$ taken at the frequency 431.8 GHz in ascending and
descending fields ($T$=4.2 K). The spectrum clearly indicates a
magnetic field hysteresis.}
\end{center}
\end{figure}

The first ESR investigation of the high-field, incommensurate
phase in CuGeO$_3$ was performed by Palme $et.~ al.$\cite{Palme},
who observed magnetic field hysteresis effects in the
incommensurate phase. Drastic changes were noted on both the ESR
linewidth and field, depending on the magnetic field sweep
direction. Generally speaking, a
    hysteresis phenomenon is a quite common feature of incommensurate
    structures\cite{Blink}, which can be explained in terms of pinning of the
    microscopical incommensurate superstructure on
    the discreteness of the crystal lattice and/or defects. If the incommensurability originates from an interplay of spin and lattice
 degrees of freedom (i.e. a magnetic structure is incommensurate with the crystallographic
 structure), the discreteness of the magnetic lattice and/or magnetic defects
    can strongly affect incommensurate
    superstructure\cite{Kiryukhin3}.

 A typical ESR spectrum at
the frequency of 431.8 GHz ($T$=4.2 K) is shown in
Fig.~\ref{Fig1}. We confirm a hysteresis behavior of the
absorption in CuGeO$_3$ in the incommensurate phase ($B>B_{DI}$).
One can see that the ESR line is much narrower in the descending
fields. Qualitatively, such a behavior can be explained as
follows. The soliton-like phase consists of nearly commensurate
regions separated by domain walls (solitons) where the phase of
the order parameter changes rapidly. Because of that, a local
field on Cu$^{2+}$ sites in CuGeO$_3$ is microscopically
modulated, that removes the equivalence of the ESR active sites
and causes spreading of the ESR absorption into a quasi-continuous
distribution of local resonance lines. Magnetic field tends to
polarize spins, making effective fields on the Cu$^{2+}$ sites
more homogenous. This results in the ESR line-narrowing, as seen
in descending fields.

\begin{figure}[!b]
\begin{center}
\vspace{2cm}
\includegraphics[width=0.45\textwidth]{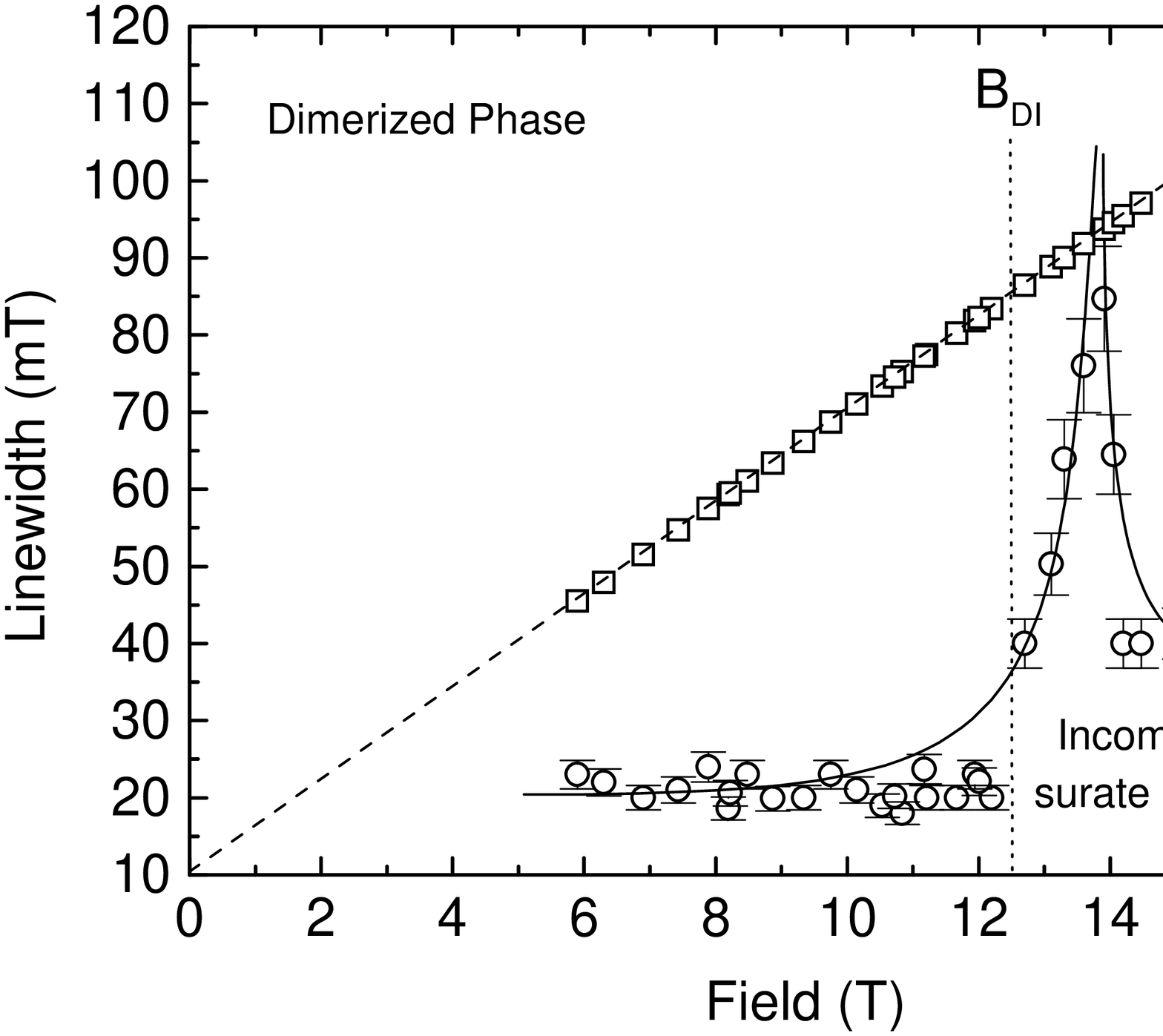}% Here is how to import EPS art
\vspace{-1.2cm} \caption{\label{Fig2} The frequency-field
(squares) and linewidth-field (circles) dependencies of the ESR
excitations in CuGeO$_3$ at $T$=4.2 K. The data are shown for
ascending fields. The dash line is a frequency-field dependence of
magnetic excitations with $g$=2.15. The solid lines are guides for
eyes. The dot line denotes the dimerized-incommensurate phase
transition boundary.}
\end{center}
\end{figure}

In Fig.~\ref{Fig2} we show a frequency and a linewidth vs magnetic
field diagrams of the ESR in CuGeO$_3$ in ascending magnetic
fields up to 17 T, and in a frequency range of 175-510 GHz.  The
Lorentzian fit of absorptions was used to calculate the ESR
linwidth at half-height. The $g$-factor of excitations remains
almost constant in the entire frequency-field range, $g\sim$ 2.15,
which is consistent with pulsed-field ESR data \cite{Nojiri2}.
However, a drastic change in the ESR linewidth $\Delta B$ is
observed at the transition from the dimerized to incommensurate
phase. A maximum in the linewidth is found at $B_{c}\sim$ 13.8 T.

 In order to explore the nature of the ESR linewidth anomaly and
 the
possible role of the soliton subsystem in it, ESR on CuGeO$_3$
+0.8\%Si (where a long-range-ordered incommensurate state appears
to be completely suppressed by doping) was performed.

It was shown that doping can significantly affect low-temperature
magnetic properties of CuGeO$_3$, creating defects and enhancing
three-dimensional antiferromagnetic correlations in the dimerized
phase\cite{Regnault}. It was found also that even a very small
doping had a drastic effect on the shape of the lattice
modulation\cite{Christianson}. The effect is especially strong in
the case of Si-doping, when Si$^{4+}$ substitutes  Ge$^{4+}$. It
distorts the lattice and the configuration of oxygens around the
copper sites, and may results in reversing coupling from
antiferromagnetic to ferromagnetic\cite{Geertsma}. If the doping
exceeds some critical concentration a long-range order in the
soliton lattice can be completely suppressed\cite{Kiryukhin}.
High-field neutron scattering experiments\cite{Grenier} revealed
only a short-range ordering of solitons in 0.7\% Si-doped samples
(while a long-range-ordered soliton structure still persists in
0.3\%Si-doped crystals), that suggests a threshold concentration
of about 0.5-0.6\%.

The doped CuGeO$_3$ samples were initially characterized by
measuring magnetic susceptibility at temperatures down to 1.8 K,
using SQUID magnetometer. The susceptibility of doped crystals
exhibits a minimum at $T\sim$ 7.7 K (an evidence of the coexisting
dimer liquid state and enhanced three-dimensional
short-range-order antiferromagnetic correlations), and a
pronounced peak, corresponding to an antiferromagnetic ordering
with $T_{N}\sim$ 3.7 K. The data are consistent with results
 obtained by Grenier $et.~al.$\cite{Grenier2} on 0.8\%Si-doped CuGeO$_3$.

\begin{figure}
\begin{center}
\vspace{2cm}
\includegraphics[width=0.45\textwidth]{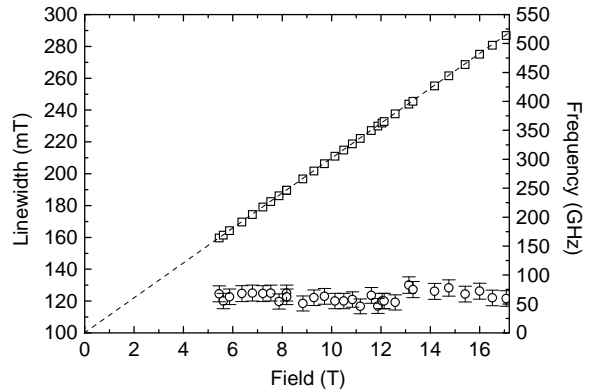}% Here is how to import EPS art
\vspace{-1.2cm} \caption{\label{Fig3} The frequency-field
(squares) and linewidth-field (circles) dependencies of the ESR
excitations in 0.8\% Si-doped CuGeO$_3$ at $T$=4.2 K. The dash
line is a frequency-field dependence of magnetic excitations with
$g$=2.15. }
\end{center}
\end{figure}

In Fig.~\ref{Fig3} we show a frequency and a linewidth vs field
diagrams of the magnetic excitations in the 0.8\%Si-doped
CuGeO$_3$ samples. Similar to the pure CuGeO$_3$, no drastic
changes are found in the $g$-factor behavior. Instead, two
distinguishing features in the ESR spectra are found. First, no
hysteresis effects are observed in fields up to 17 T, which
appears to be an evidence of the collapsing long-range ordered
soliton-like lattice. Second, the $\Delta B$ anomaly found in the
pure CuGeO$_3$ at $B_{c}\sim$ 13.8 T, is completely suppressed in
 doped CuGeO$_3$.

Our observations clearly indicate the essential role of the
long-range-order soliton correlations in the ESR linewidth anomaly
in CuGeO$_3$. Like any structural imperfection in spin systems
with a collective type of elementary excitations (note for
instance that the ESR linewidth in the dimerized phase in pure
CuGeO$_3$ is about six times smaller than that in the doped
samples, Fig.~\ref{Fig2} and Fig.~\ref{Fig3}), the soliton lattice
in CuGeO$_3$ introduces additional scattering for magnons. As a
result, an intensive magnon-soliton scattering manifests itself in
the ESR line-broadening. A maximum of the linewidth is observed at
$B_{c}\sim$ 13.8 T, that clearly indicates a pronounced
development of the incommensurate soliton-like superstructure (and
a corresponding enhancement of the scattering processes) close to
the boundary of the dimerized-incommensurate phase transition,
$B_{DI}$. This observation is consistent with high-field
magnetostriction and thermal expansion experiments\cite{Lorenz}.

In conclusion, the field-induced structural evolution in the
spin-Peierls compound CuGeO$_3$ is probed using frequency-tunable
high-resolution ESR, in fields up to 17 T. Our studies reveal
several important peculiarities of its high-field properties.  The
ESR linewidth anomaly strongly suggests the essential role of
magnon-soliton scattering processes in the soliton-like phase, and
confirms that the soliton-like regime exists close to the boundary
of the dimerized-incommensurate phase transition. Our data are
consistent with high-field inelastic neutron scattering
experiments, suggesting that doping significantly affects the
solitonlike structure in CuGeO$_3$, suppressing long-range-order
soliton correlations and corresponding magnon-soliton scattering.
The use of the high-field frequency-tunable ESR approach (applied
for an analysis of the ESR linewidth in a broad frequency-field
range) can provide important information  on field-induced
structural evolutions in other spin-Peierls materials\cite{Bray}.

Acknowledgement.  The authors would like to thank S.~McCall and
Z.~Zhou for performing magnetic susceptibility measurements. The
25 T resistive magnet was built with financial
    support of the W.M.~Keck Foundation of Los Angeles, CA.

\end{document}